\documentclass[10pt,conference,letterpaper,final]{IEEEtran}
\IEEEoverridecommandlockouts

\usepackage{graphicx}
\usepackage[table]{xcolor}
\usepackage{tikz}
\usepackage{amsmath}
\usepackage{amssymb}
\usepackage{amsfonts} 
\usepackage[normalem]{ulem}
\usepackage{bbm}
\usepackage{url}
\usepackage{makecell} 
\usepackage{xstring}

\usepackage{algorithm}
\usepackage[noend]{algpseudocode}
\algrenewcommand\algorithmicrequire{\textbf{Input:}}
\algrenewcommand\algorithmicensure{\textbf{Output:}}
\algrenewcommand{\algorithmiccomment}[1]{\hfill{\color{blue}$\triangleright$#1}}

\usepackage{siunitx}
\usepackage{mathtools,zref-savepos}

\usepackage{float}
\usepackage[caption=false,font=footnotesize,justification=centering,format=hang]{subfig}
\usepackage{booktabs}  
\usepackage{tabularx}  
\usepackage{multirow}
\usepackage{xspace}

\usepackage{ifdraft}
\ifoptionfinal{\usepackage[final]{fixme}}{\usepackage[draft]{fixme}}
\fxsetup{mode=multiuser,theme=color,layout=inline,nomargin} 
\FXRegisterAuthor{DL}{dl}{\color{teal}\fbox{DL}}
\FXRegisterAuthor{DB}{db}{\color{blue}\fbox{DB}}
\FXRegisterAuthor{OK}{ok}{\color{olive}\fbox{OK}}
\FXRegisterAuthor{GY}{gy}{\color{purple}\fbox{GY}}

\newcommand{\DL}[1]{\DLnote{\color{teal}#1}}
\newcommand{\DB}[1]{\DBnote{\color{blue}#1}}

\newcommand{\GY}[1]{\GYnote{\color{purple}#1}}

\usepackage{balance}

\usepackage{xspace}
\usepackage{xparse}
\usepackage{enumitem}

\usepackage{comment}

\usepackage[hidelinks]{hyperref}

\usepackage[capitalise]{cleveref}
\crefname{equation}{}{}
\crefname{section}{Sec.}{Secs.}

\newcommand{\Set}[1]{\ensuremath{\mathcal{#1}}\xspace}
\newcommand{\Shortstack}[2][t]{\begin{tabular}[#1]{@{}l@{}}#2\end{tabular}}

\newcommand{\bluetext}[1]{\textcolor{blue}{#1}}

\DeclareSIUnit{\million}{\text{million}}

\crefname{define}{Definition}{Definitions}

\crefname{observation}{Observation}{Observations}
\newtheorem{lemma}{Lemma}
\newtheorem{theorem}{Theorem}

\newcommand{\dc}{datacenter\xspace}
\newcommand{\dcs}{datacenters\xspace}

\newcommand{\Arnes}{Arnes-SI\xspace}
\newcommand{\Amres}{Amres-RS\xspace}
\newcommand{\Dfn}{Dfn-DE\xspace}
\newcommand{\NTU}{node-TU\xspace}
\newcommand{\LTU}{link-TU\xspace}

\newcommand\MyIncludeGraphics[2][]{
    \IfFileExists{#2}{%
        \includegraphics[#1]{#2}%
    }{%
        \missingfigure[figwidth=7.0cm]{Missing #2}%
    }%
}%

\newcommand{\T}[1]{\par\noindent\rule{0pt}{\baselineskip}\textit{#1:}}

\newcommand{\set}[1]{\left\{#1\right\}}         

\newcommand{\abs}[1]{\left\vert#1\right\vert}

\DeclareMathOperator*{\argmin}{arg\,min}  

\DeclareMathOperator{\concat}{+\kern -0.4em+}

\newcommand{\newVar}[2]{\newcommand{#1}{\ensuremath{#2}\xspace}}
  \newVar{\server}{S}
  \newVar{\client}{C}
  \newVar{\rclient}{R_c}
  \newVar{\rserver}{R_s}

\providecommand{\ie}{\emph{i.e.,} }

\newcommand{\ignore}[1]{}
\newcommand{\Small}{\footnotesize} 

\setlist[description]{leftmargin=*}
\usepackage{float}
\floatstyle{ruled}

\newcommand{\vnep}{\textsc{vnep}\xspace}
\newcommand{\vneap}{\textsc{vneap}\xspace}
\newcommand{\Vneap}{Virtual Network Embedding with Alternatives Problem\xspace}
\Crefname{problem}{Problem}{Problem}
\crefname{prob}{Problem}{Problem}
\Crefname{prob}{Problem}{Problem}
\crefformat{prob}{#2\vneap{}#3}
\Crefformat{prob}{\Vneap}

\crefname{equation}{Eq.}{Eqs.}
\Crefname{equation}{Eq.}{Eqs.}
\crefname{figure}{Fig.}{Figs.}
\Crefname{figure}{Fig.}{Figs.}
\crefname{alt}{Alternative}{Alternatives}
\Crefname{emb}{Embedding}{Embeddings}
\crefname{emb}{}{Embeddings}
\crefname{greedy}{}{}
\creflabelformat{greedy}{#2\textsc{\textbf{G\scalebox{.8}{\kern-1pt reedy}}}#3}
\crefname{tanto}{}{}
\creflabelformat{tanto}{#2\textsc{\textbf{T\scalebox{.8}{\!anto}}}#3}
\crefname{embed}{}{}
\creflabelformat{embed}{#2\textbf{Embed}#3}
\crefname{algorithm}{Alg.}{Algs.}

\newlist{doforall}{enumerate*}{1}
\setlist[doforall]{label=\textsuperscript{\alph*}}

\newfloat{problem}{tbp}{lop}
\floatname{problem}{Problem}

\newcommand{\blue}[2][blue]{{\color{#1}#2}}
\newcommand{\Rplus}{\mathbb{R}^+}  

\newcommand{\myop}[1]{\operatorname{#1}}

\newcommand{\size}[1]{\myop{load}(#1)}
\newcommand{\capacity}[1]{\myop{cap}(#1)}
\newcommand{\cost}[1]{\myop{cost}(#1)}

\newcommand{\netsym}{\!s}
\newcommand{\graphsym}{\boldsymbol{G}}
\NewDocumentCommand\gnet{d<>}
    {\IfNoValueTF{#1}
        {\graphsym_{\netsym}}
        {\graphsym_{\netsym}(#1)}}
\newcommand{\vnet}{V_{\!\netsym}}
\newcommand{\enet}{E_{\netsym}}
\NewDocumentCommand\pnet{o}{\boldsymbol{p}\IfNoValueTF{#1}{}{_{#1}}}
\newcommand{\lnet}[1][vw]{(#1)}
\NewDocumentCommand\gapp{d<>O{t}}
    {\IfNoValueTF{#1}
        {\graphsym_{#2}}
        {\graphsym(#1)_{#2}}}
\newcommand{\vapp}[1][t]{V_{\!#1}}
\newcommand{\eapp}[1][t]{E_{#1}}
\NewDocumentCommand\napp{d<>O{i}}
    {{#2}\IfNoValueTF{#1}
        {}
        {_{#1}}}
\NewDocumentCommand\lapp{d<>O{ij}}
    {(#2)\IfNoValueTF{#1}
        {}
        {_{#1}}}
\newcommand{\allapps}{\Set A}
\newcommand{\app}{a}
\newcommand{\tapp}[1][\app]{T\kern -0.3em\left(#1\right)}  
\newcommand{\Tapp}[1][\allapps]{T(#1)}  
\newcommand{\user}{\theta}   

\NewDocumentCommand\reqs{o}
    {\IfNoValueTF{#1}
        {\Set R}
        {\Set R_{#1}}}
\NewDocumentCommand\rreqs{o}    
    {\IfNoValueTF{#1}
        {\tilde{\Set R}}
        {\tilde{\Set R_{#1}}}}

\newcommand{\req}{r}
\newcommand{\rreq}{\tilde{r}}   

\newcommand{\oreq}{v(\req)}    
\newcommand{\areq}{\app(\req)}  
\newcommand{\treq}{\tapp[\areq]}    

\NewDocumentCommand\greq{d<>}
    {\IfNoValueTF{#1}
        {\graphsym(\req)}
        {\graphsym(\req,#1)}}
\NewDocumentCommand\nreq{d<>O{i}}
    {\IfNoValueTF{#1}
        {\napp<\req>[#2]}
        {\napp<\req,#1>[#2]}}
\NewDocumentCommand\lreq{d<>O{ij}}
    {\IfNoValueTF{#1}
        {\lapp<\req>[#2]}
        {\lapp<\req,#1>[#2]}}

\NewDocumentCommand\Xreq{o}
    {\IfNoValueTF{#1}
        {\boldsymbol{x}}
        {\boldsymbol{x}_{#1}}}
\NewDocumentCommand\rXreq{o} 
    {\IfNoValueTF{#1}
        {\boldsymbol{\tilde{x}}}
        {\boldsymbol{\tilde{x}}_{#1}}}

\NewDocumentCommand\xreq{d<>O{\req}mm}
    {\IfNoValueTF{#1}
        {x^{#3}_{#4}(#2)}
        {x^{#3,#1}_{#4}(#2)}}

\NewDocumentCommand\rxreq{d<>O{\req}mm}  
    {\IfNoValueTF{#1}
        {\tilde{x}^{#3}_{#4}(#2)}
        {\tilde{x}^{#3,#1}_{#4}(#2)}}

\NewDocumentCommand\Yreq{o}
    {\IfNoValueTF{#1}
        {\boldsymbol{\tilde{y}}}
        {\boldsymbol{\tilde{y}}_{#1}}}
\NewDocumentCommand\yreq{d<>O{\rreq}mm}  
    {\IfNoValueTF{#1}
        {\tilde{y}^{#3}_{#4}(#2)}
        {\tilde{y}^{#3,#1}_{#4}(#2)}}

\newcommand{\Xopt}{\Xreq^{\textsc{v{\tiny\!neap}l\tiny{p}}}}
\newcommand{\rXopt}{\rXreq^{\textsc{v{\tiny\!neap}l\tiny{p}}}}
\newcommand{\Xtant}{\Xreq^{\textsc{t\tiny\!anto}}}

\newcommand{\dreq}{d(\req)}
\newcommand{\rdreq}{\tilde{d}_{\req}}

\newcommand{\dapp}[1]{{\beta}^{#1}}
\newcommand{\dnet}[2]{{\eta}^{#1}_{#2}}

\NewDocumentCommand\vnea{d()}
    {\ensuremath{\IfNoValueTF{#1}
        {\operatorname{VNEA}}
        {\operatorname{VNEA}(#1)}}}

\newcommand{\embedreject}{Embed-or-Reject}

\newcommand{\WRS}[1]{\operatorname{\textsc{wrs}}(#1)}

\newcommand{\OPT}{\textsc{\textbf{V\scalebox{.8}{\!neap}L\scalebox{.8}{p}}}\xspace}
\newcommand{\aOPT}{\textsc{\textbf{V\scalebox{.8}{\!nep($\gapp[1]$)}}}\xspace}
\newcommand{\bOPT}{\textsc{\textbf{V\scalebox{.8}{\!nep($\gapp[2]$)}}}\xspace}

\newcommand{\Greedy}{\cref{greedy}\xspace}
\newcommand{\aGreedy}{\textsc{\textbf{S\scalebox{.8}{a1}G\scalebox{.8}{\kern-1pt reedy}}}\xspace}
\newcommand{\bGreedy}{\textsc{\textbf{S\scalebox{.8}{a2}G\scalebox{.8}{\kern-1pt reedy}}}\xspace}
\newcommand{\VNEAP}{\cref{prob:vnea}\xspace}
\newcommand{\TANTO}{\cref{tanto}\xspace}

\usepackage{enumitem}
\usepackage{eso-pic}
\begin{document}
\AddToShipoutPictureBG*{%
  \AtTextLowerLeft{%
    \raisebox{-2\height}{%
      \parbox{\textwidth}{\centering%
      \copyright 2025 IEEE. Personal use of this material is permitted.\\ 
      Permission from IEEE must be obtained for all other uses, in any current or future media, including reprinting/republishing this material for advertising or promotional purposes, creating new collective works, for resale or redistribution to servers or lists, or reuse of any copyrighted component of this work in other works.} 
      \hspace{\columnsep}\makebox[\columnwidth]{}
    }%
  }%
}

\title{The Power of Alternatives in Network Embedding}


\author{
    \IEEEauthorblockN{Oleg Kolosov}
    \IEEEauthorblockA{Computer Science \\
    Technion\\
    Israel \\
    \href{mailto:kolosov@campus.technion.ac.il}{kolosov@campus.technion.ac.il} 
    }
\and
    \IEEEauthorblockN{Gala Yadgar}
    \IEEEauthorblockA{Computer Science \\
    Technion\\
    Haifa, Israel \\
    \href{mailto:gala@cs.technion.ac.il}{gala@cs.technion.ac.il}
    }
\and
    \IEEEauthorblockN{David Breitgand}
    \IEEEauthorblockA{Hybrid Cloud \\ 
    IBM Research -- Haifa \\
    Haifa, Israel \\
    \href{mailto:davidbr@il.ibm.com}{davidbr@il.ibm.com}
    }
\and
    \IEEEauthorblockN{Dean H. Lorenz}
    \IEEEauthorblockA{Hybrid Cloud \\ 
    IBM Research -- Haifa \\
    Haifa, Israel \\
    \href{mailto:dean@il.ibm.com}{dean@il.ibm.com}
    }
\and
    \IEEEauthorblockN{Rasoul Behravesh}
    \IEEEauthorblockA{\textit{6G Research}\\
    \textit{Samsung R\&D Institute}\\
    Surrey, United Kingdom\\
    \href{mailto:r.behravesh@samsung.com}{r.behravesh@samsung.com}
    \thanks{Part of this work were conducted while Rasoul Behravesh was affiliated with Fondazione Bruno Kessler, Trento, Italy.}
    }
}

\author{
    \IEEEauthorblockN{%
    Oleg Kolosov\IEEEauthorrefmark{1},
    Gala Yadgar\IEEEauthorrefmark{1}, 
    Rasoul Behravesh\IEEEauthorrefmark{3},
    David Breitgand\IEEEauthorrefmark{2},
    and Dean H. Lorenz\IEEEauthorrefmark{2}%
    }
    \IEEEauthorblockA{
    \IEEEauthorrefmark{1}
    \textit{Computer Science, Technion}, Israel.
    \{\href{mailto:kolosov@cs.technion.ac.il}{kolosov},\href{mailto:gala@cs.technion.ac.il}{gala}\}@cs.technion.ac.il 
    }
    \IEEEauthorblockA{
    \IEEEauthorrefmark{3}
    \textit{6G Research, Samsung R\&D Institute}, United Kingdom.
    \href{mailto:r.behravesh@samsung.com}{r.behravesh@samsung.com}
    \thanks{Part of this work were conducted while Rasoul Behravesh was affiliated with Fondazione Bruno Kessler, Trento, Italy.
    This work was partially funded by the European Union’s Horizon 2020 research and innovation program, the IBM–Technion Research Collaboration, US-Israel BSF grant 2021613, and ISF grant 807/20.
    }
    }
    \IEEEauthorblockA{
    \IEEEauthorrefmark{2}
    \textit{Hybrid Cloud, IBM Research -- Israel}, Israel.
    \{\href{mailto:davidbr@il.ibm.com}{davidbr},\href{mailto:dean@il.ibm.com}{dean}\}@il.ibm.com
    }
}

\maketitle

\begin{abstract}
In the \textit{virtual network embedding problem}, the goal is to map (\textit{embed}) a set of virtual network instances to a given physical network substrate at minimal cost, while respecting the capacity constraints of the physical network. This NP-hard problem is fundamental to network virtualization, embodying essential properties of resource allocation problems faced by service providers in the edge-to-cloud spectrum. Due to its centrality, this problem and its variants have been extensively studied and remain in the focus of the research community. 

In this paper, we present a new variant, the \textit{virtual network embedding with alternatives problem (\VNEAP)}. This new problem captures the power of a common network virtualization practice, in which virtual network topologies are malleable---embedding of a given virtual network instance can be performed using any of the alternatives from a given set of topology alternatives. We provide two efficient heuristics for \vneap and show that having multiple virtual network alternatives for the same application is superior to the best results known for the classic formulation.  We conclude that capturing the problem domain via \vneap can facilitate more efficient network virtualization solutions.

\end{abstract}

\setlist[itemize]{leftmargin=*}
\def\sscale{.8}



\section{Introduction}\label{sec:into}

Network virtualization reshapes the way networked applications are provisioned across the edge-cloud continuum. This technology enables networked applications to be defined entirely in software as \textit{virtual networks}, where nodes represent \textit{virtual network functions} (VNFs), and virtual links specify communication requirements between these VNFs. Specific virtual network instances are provisioned on demand in response to customer requests for particular applications, while the underlying physical network substrate continues to be used as the packet-forwarding data plane. 

To leverage the benefits of network virtualization, the application provider needs to solve the \textit{virtual network embedding problem (\vnep)} whose input is a set of requests to provision application instances, represented by their virtual \textit{topologies}. \textit{Embedding} requires mapping each virtual node to a physical substrate node and each virtual link to a physical substrate path. The objective is to handle as many requests as possible while minimizing cost or maximizing profit (\ie gains obtained for successful embeddings~\cite{rost2019virtual}).
%
\vnep is known to be strongly NP-hard~\cite{Rexford2008,RostSchmidt-ToN2020}. 
Due to \vnep's fundamental importance, efficient sub-optimal solutions have been extensively studied~\cite{Rexford2008,RostSchmidt-ToN2020,behravesh2024practical,chowdhury2009virtual,RostDohneSchmid-parameterized-2019,rost2019virtual,feng2017approximation,SCHARDONG2021107726}.

In practice, the efficiency of a heuristic depends on how well it leverages domain properties. In this work, we propose exploiting a common practice in network virtualization that is not captured by the classic \vnep formulation. Instead of considering a single virtual network per application provisioning request, we allow selection from a set of \textit{alternatives}. These alternatives are functionally equivalent but are implemented as different virtual networks with different capacity and bandwidth requirements. This aligns with the well-established practice in network virtualization of enabling flexible configurations.
It allows better adaptation to non-uniformity in application request distributions and in the physical substrate. For example, in the edge-to-cloud spectrum, \dcs are heterogeneous: \dcs closer to the central cloud have larger capacities, cheaper resources, and higher bandwidth interconnect. The smaller edge \dcs have more expensive resources and lower bandwidth.

Imagine embedding an application instance with two alternative topologies. The first topology comprises two virtual nodes with small computational requirements but substantial bandwidth requirements. The second topology includes a compression function, which increases the computational requirements but significantly reduces the required bandwidth. The first topology is ideal when compute costs are high and communication costs are low. The second topology is preferable when compute costs are low and communication costs are high. In the classic \vnep, only a single topology is considered. However, in practice, multiple alternative configurations of the same virtual network are often prepared to better adapt to actual deployment conditions. 
Thus, a new problem formulation is required to enable consideration of both alternatives per each application embedding request. For each request, the most cost-efficient embedding of one alternative should be found, such that the cost of embedding all requests is minimized while aiming at a minimal rejection rate (\ie minimizing the number of requests for which an embedding could not be found). 

To this end, we define a new problem: the \textit{\Vneap (\vneap)}. \vneap receives 
 (1) a physical network substrate with costs of hosting VNFs in the \dcs, capacities of the \dcs and their interconnecting physical links, and the cost of communication over each link; 
 (2) a set of applications, each comprising a set of \textit{alternative topologies},  
where each alternative is characterized by VNF capacity requirements and bandwidth requirements of inter-VNF links;  
(3) a set of application embedding requests. 
The goal is to embed a maximum number of application requests into the substrate, at a minimized cost, by selecting an optimized \textit{combination} of alternatives. 
We study \vneap and propose two scalable heuristics: (a) \Greedy and (b) \textit{\textbf{T}ree of \textbf{A}lternative \textbf{N}etwork \textbf{TO}pologies (\cref{tanto})}, a scalable near-optimal global optimization. 

Our key research goal is to determine whether allowing a mix of alternative topologies for the same application yields substantial cost-efficiency gains over the classical single-alternative \vnep. Our secondary goal is to characterize the scenarios where global optimization outperforms a greedy strategy. To this end, we conducted a large-scale experimental study using substrate and application topologies previously reported in the literature. Our results clearly show that using alternatives is a superior strategy. They also demonstrate that even a small number of alternatives can result in significant improvement in the system's cost-efficiency and utilization. 

In summary, we make the following contributions. We identify the limitations of \vnep (\cref{sec:toy}) and define \VNEAP (\cref{sec:problem}).  We propose \Greedy and \cref{tanto} and provide a bound on the gap between \cref{tanto} and the optimal solution, which we prove to be independent of the number of requests (\cref{sec:solution}). We demonstrate that \vneap facilitates embeddings with significantly higher cost efficiency and utilization than classic \vnep (\cref{sec:evaluation}). We show that \cref{tanto} global optimization is significantly superior to greedy heuristics and its results are very close to the theoretical fractional optimum.

\section{Related Work}\label{sec:related}
Selecting from multiple functionally equivalent application topologies is a known technique in network virtualization~\cite{xavier2017elastic,konstantoudakis2021serverless}. However,
to the best of our knowledge, this work is the first to formally define and systematically study \cref{prob:vnea}, which is a generalization of \vnep. A subtle but crucial difference is that alternative topologies, eligible for the same application embedding request, cannot be regarded as separate requests. Rather \cref{prob:vnea} seeks the most cost-efficient combination of topological alternatives, assigning each request a unique alternative, and embedding the request using this selection.

A recent study shows that \vnep is NP-complete 
and its optimization variants are NP-hard. 
Furthermore, \vnep remains NP-hard even if only node placement and edge routing restrictions are considered~\cite{RostSchmidt-ToN2020}. \VNEAP is at least as complex as \vnep, as it stems from \vnep.
Given \vnep complexity and significance, efficient heuristics are of great importance~\cite{rost2019virtual} and this problem remains in the focus of research efforts. A classic approach is to solve a relaxed LP formulation and then round a solution.
Our rounding techniques have similar properties to the randomized rounding of~\cite{rost2019virtual} and lead to provable optimality gaps under a similar assumption, namely, that a single request is much smaller than the substrate node and link capacities. 

\vnep is actively studied in the context of \textit{service-function chain (SFC)} deployment, where embedding a \textit{VNF forwarding graph} is equivalent to solving \vnep. Approaches to solving \vnep include greedy heuristics~\cite{BehraveshHCR-TNSM2021,ChowdhuryRB-TON2012,kolosov2023pase}, metaheuristics~\cite{Ruiz2018, Cao2017, Kiran2021}, and reinforcement learning techniques~\cite{NFVdeep-2019, DeepRL-2021}.
In SFC deployment, trivial alternatives are considered in the form of an inventory of different VNF \textit{flavors} (\ie sizes). However, alternative topologies with entirely different functions are not part of the forwarding graph definition. Thus, current solutions to \vnep do not apply in the general case. To the best of our knowledge, our work is the first to facilitate the optimal combination of different but functionally equivalent topologies for virtual network embedding. 

A scalable heuristic for \vnep was recently proposed for SFC deployment~\cite{behravesh2024practical}. This heuristic operates under assumptions on the physical substrate and application topologies (SFC chains) similar to those studied in this work. It also addresses latency requirements and achieves near-optimal results. It inspired the design of \cref{tanto}, which shares similar properties when limited to one topology per application. However, there is a fundamental difference between these works. While~\cite{behravesh2024practical} focuses primarily on scalability, our work centers on the novel optimization problem we define and explores its expressive power and impact on achievable cost and resource utilization.
 


\begin{figure}
    \centering
    \begin{minipage}[b]{0.48\linewidth}\centering
        \includegraphics[page=2,trim = 0cm .1cm 0cm 0cm]{fig/FluidModelTikz.pdf}
    \caption{Alternative topologies\label{fig:alttop}}
    \end{minipage}
    \hfil
    \begin{minipage}[b]{0.48\linewidth}\centering
        \includegraphics[page=3,trim = 0cm 0cm 0cm -0.2cm]{fig/FluidModelTikz.pdf}
    \caption{Substrate topology\label{fig:substrate}}
    \end{minipage}
    \vspace{-4mm}
\end{figure}

\section{A Simple Example}
\label{sec:toy}
To build an intuition, we provide a straightforward example to illustrate the tradeoffs inherent in \VNEAP.
\Cref{fig:alttop} represents an application with two virtual topology alternatives. The main alternative (solid lines) has two interconnected VNFs, $A$ of size $5$ and $B$ of size $100$. The alternative topology (dotted red lines) adds an accelerator VNF ($acc$) of size $10$. While it increases the compute requirements, it reduces the bandwidth requirements into VNF $B$ from $100$ to $30$. 
We embed this application onto the substrate shown in \cref{fig:substrate}, comprising an Edge \dc connected to a Core \dc, with given capacities and costs. In this simple case, the Core \dc is 10 times cheaper and 10 times larger than the Edge \dc.  
\begin{figure}
\def\sscale{1.0}
    \centering\hfill
    \subfloat[][\label{fig:example1}]{\label[emb]{e1}
        \includegraphics[page=7,scale=\sscale]{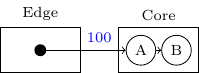}
    }\hfill
    \subfloat[][\label{fig:example2}]{\label[emb]{e2}
        \includegraphics[page=4,scale=\sscale]{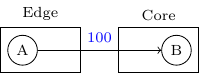}
    }\hfill\mbox{}\\[-1.5\baselineskip]\hfill
    \subfloat[][\label{fig:example3}]{\label[emb]{e3}
        \includegraphics[page=6,scale=\sscale]{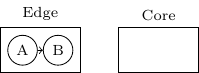}
    }\hfill
    \subfloat[][\label{fig:example4}]{\label[emb]{e4}
        \includegraphics[page=5,scale=\sscale]{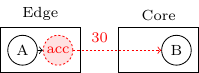}
    }\hfill\mbox{}
    \caption{Embedding with alternatives\label{fig:altembed}}
    \vspace{-1em}
\end{figure}

\Cref{fig:altembed} depicts all reasonable embedding options for a request made at the Edge \dc. \Cref{fig:example1,fig:example2,fig:example3} show different embeddings of the main topology, while \cref{fig:example4} shows an embedding of the alternative topology.
\Cref{e1} incurs the lowest compute cost ($105$) by placing both VNFs on the cheaper Core \dc. However, it incurs the highest bandwidth cost ($100$).  
\Cref{e2} places $A$ on the Edge \dc and $B$ on the Core \dc, incurring the same bandwidth cost ($100$) as \cref{e1} but a higher compute cost ($150$). 
\Cref{e2} might be chosen due to other constraints, such as VNF $A$ being limited to the Edge. 
\Cref{e3} has no bandwidth cost as $A$ and $B$ are collocated on the Edge \dc, but it incurs the highest compute cost ($1050$), since it uses the more expensive Edge \dc.
\Cref{e4} is the only reasonable option for utilizing the alternative topology. It is similar to \cref{e2}, but uses the extra accelerator VNF to reduce bandwidth cost. The compute cost increases to $250$, but the bandwidth cost is only $30$. 

The optimal embedding for an incoming request depends on available compute and bandwidth resources and on the relative compute-to-bandwidth cost. With the given substrate costs, \cref{e1} is the cheapest option in a lightly loaded system. If there are many requests, the Core \dc may become saturated. An optimal embedding of \emph{all} requests may need to use \cref{e2} to offload some of the compute resources to the Edge. In a highly loaded system, the core may be so overloaded, to the point that the only available embedding would be the expensive \cref{e3}. 

The alternative virtual topology would come into play when the links become overloaded. For example, suppose the substrate link capacity was only $5K$, instead of $10K$, and there are $100$ requests. Without the alternative topology, only $50$ requests could be served with \cref{e1} before saturating the link. Therefore, $50$ requests would be served by the very expensive \cref{e3}. An optimal solution in this case would utilize the alternative topology to serve $100$ requests with \cref{e4}, reducing more than half of the overall cost. Another example where \cref{e4} is preferred is if bandwidth costs are high (e.g., if the substrate link would cost $3$ instead of $1$). 

\section{Virtual Network Embedding with Alternatives Problem}\label{sec:problem}

\begin{table}
    \caption{Notations}
    \label{tab:notations}
    \centering\setlength{\tabcolsep}{2pt}
    \scriptsize
    \begin{tabularx}{\columnwidth}{@{}lXr@{}}
        \toprule
        \textbf{Notation} & \textbf{Description} & \\ 
        \midrule
        $\gnet(\vnet, \enet)$ & substrate graph 
        & \multirow[b]{1}{*}{\llap{\color{blue}\shortstack[r]{Substrate}}} \\
        $\cost{v}, \cost{vw}$ & cost of resources on node $v$, link $\lnet$ \\
        $\capacity{v}, \capacity{vw}$ & available resources on node $v$, link $\lnet$\\
        $\size{v}, \size{vw}$ & load induced on node $v$, link $\lnet$\\
        \midrule
        $a \in \allapps$ & an application with alternatives
        & \multirow[b]{1}{*}{\llap{\color{blue}\shortstack[r]{Applications}}} \\
        $\tapp = \set{ \gapp[1], \gapp[2], \ldots, \gapp, \ldots}$ & alternative virtual network topologies of app. $a$ \\
        $\napp<t>, \lapp<t>$ & virtual node, link of topology $\gapp$ \\
        $\dapp{i}, \dapp{ij}$ & size of virtual node $i$, link $\lapp$\\
        $\dnet{i}{v}, \dnet{ij}{vw}$ & (in)efficiency of serving  $i$ on $v$, $\lapp$ on $\lnet$\\
        \midrule
        $\req = ( \oreq, \areq, \dreq) \in \reqs$ & a VNEA request (origin, app, size) 
        & \llap{\color{blue}Requests} \\
        $\reqs[ua] \subseteq\reqs$ & similar requests: $\set{\req : u=\oreq \land a=\areq}$ \\
        $\Psi, \psi$ & rejection penalty: $\psi d_r$ is the cost of rejecting $r$ \\
        \midrule
        $\greq \in \treq$ & topology chosen for $\req$
        & \multirow[b]{1}{*}{\llap{\color{blue}\shortstack[r]{Embedding}}} \\
        $\xreq<t>{i}{v}, \xreq<t>{ij}{vw} \in \set{0,1}$ & $1$ iff $\napp<t>$ mapped to $v$, $\lapp<t>$ to $\lnet$ for $\req$\\
        \bottomrule
    \end{tabularx}
    \vspace{-3mm}
\end{table}

\cref{prob:vnea} shares many common definitions and notations with \vnep. Therefore, we first recapitulate the definitions and notations common to both \cref{prob:vnea} and \vnep (for completeness) in~\cref{subsec:preliminaries}. Next, we provide definitions and notations unique to \cref{prob:vnea} in~\cref{subsec:preliminaries-vneap}. In~\cref{subsec:probstat} we provide a formal problem formulation in the form of MILP.

\subsection{Definitions and Notations Common to \vneap and \vnep}\label{subsec:preliminaries}

\T{Physical substrate network} A set of interconnected \dcs. We model it as a graph $\gnet(\vnet, \enet)$, where nodes represent \dcs and links represent connections between them. The functions $\cost{}$ and $\capacity{}$ define the resource usage cost and capacity for each node $v \in \gnet$ and each link $\lnet \in \gnet$.

\T{Applications} Given an application set $\allapps$, each application $\app \in \allapps$ is defined by its virtual network topology graph, $\gapp[a](\vapp[a], \eapp[a])$. A virtual node $\napp \in \gapp[a]$ represents a VNF $i$ in $a$; a virtual link $\lapp$ represents the interaction between VNFs $i,j$. For simplicity, we assume each virtual network topology in $\allapps$ is a tree (or chain) with $\user$ as its \emph{root} node. This assumption can be relaxed, allowing application topologies to be modeled as general graphs, by applying the techniques described in~\cite{RostDohneSchmid-parameterized-2019}.

\T{Requests} Each request $\req \in \reqs$ is modeled as a tuple $\req = ( \oreq, \areq, \dreq)$. The node $\oreq \in \gnet$ is the request's origin \dc and where the root $\user$ of its virtual network must be placed; $\areq \in \allapps$ is the requested application, which means one of $\areq$ alternative topologies must be deployed; and $\dreq \in \Rplus$ is the request's demand size, \ie the amount of resources required to embed the request. We assume requests are fully isolated, \ie they do not share VNF instances. This is a typical practical use case due to governance, compliance, security, and performance requirements. 

\T{Virtual node and link sizes (flavors)} The amount of resources required to satisfy a request for an application depends on the topology used and where it is embedded. We denote by $\dapp{i}$, the \emph{size} of a virtual node (\ie VNF) $i$.\ignore{ when supporting a single application share.} Similarly, we denote by $\dapp{ij}$ the size of a virtual link. \ignore{ when supporting a single application share.} The sizes $\dapp{i}, \dapp{ij}$ allow us to model different application topologies with different resource requirements on their nodes and links, \ie different VNF flavors. Lastly, the virtual root node, $\user$, only represents the request's location, therefore, we set $\dapp{\user} = 0$.

\T{Resource requirements}  We denote by $\dnet{i}{v}, \dnet{ij}{vw}$ the \emph{(in)efficiency} coefficient for serving $i$ on substrate node $v$ and $\lapp$ on $\lnet$, respectively. It allows us to factor in placement policies.
Using $\dnet{i}{v}$, we can model preferred substrate nodes for specific functions. A lower $\dnet{i}{v}$ means that it is preferable to place $i$ on $v$; for example, placing a packet processing function on a node with hardware acceleration support. A higher value means $i$ takes up more resources on $v$ and an extremely high value may be used to indicate that $i$ should never be mapped to $v$. Similarly,  $\dnet{ij}{vw}$, allows us to model preferred substrate links for specific traffic, e.g., to prevent certain virtual links from being mapped on some paths in the substrate networks when logical links are mapped on paths in the physical substrate network.

\subsection{Definitions and Notations Unique to \vneap}\label{subsec:preliminaries-vneap}
\T{Applications with alternatives} We model the alternative topologies of each application $\app \in \allapps$ 
as a set of graphs $\tapp = \set{ \gapp[1](\vapp[1], \eapp[1]), \gapp[2](\vapp[2], \eapp[2]), \ldots, \gapp[t](\vapp[t], \eapp[t]), \ldots }$.
The application developer provides $\tapp$ at design time to adapt to different deployment conditions. Creating $\tapp$ relies on domain-specific knowledge and is out of this work's scope.

\T{Alternative topology selection} For each request $\req \in \reqs$, select a topology $\greq \in \treq$  or \emph{reject} the request. If rejected (e.g., due to insufficient capacity), no alternative is selected for this request, and this is denoted by assigning $\greq \gets \emptyset$. 

\T{Alternative topology  embedding} Unless $\req$ is rejected ($\greq = \emptyset$), provide an embedding of $\greq \mapsto \gnet$, where nodes of $\greq$ are mapped onto nodes of $\gnet$ and links of $\greq$ are mapped onto paths in $\gnet$. The binary variable $\xreq<t>{i}{v}$ indicates whether node $\napp<t> \in \gapp$ is embedded on substrate node $v$. Similarly, the binary variable $\xreq<t>{ij}{vw}$ indicates whether link $\lapp<t> \in \gapp$ is embedded on substrate link $\lnet$. If $\gapp \neq \greq$ (or $\greq = \emptyset$) then $\xreq<t>{i}{v} \gets 0$, $\xreq<t>{ij}{vw} \gets 0$ for all $\gapp$ and $\gnet$ elements.

\subsection{Problem Statement}\label{subsec:probstat}

\Cref{fig:vnea} presents a formal mixed-integer linear program (MILP) formulation for the virtual network embedding with alternatives problem (\cref{prob:vnea}).

\begin{figure}[H]
\footnotesize\phantomsection\label[prob]{prob:vnea}
\par\noindent\rule[-.1\baselineskip]{\linewidth}{0.5pt}
\par{\textbf{\Vneap (\vneap)}}
\par\rule[.7\baselineskip]{\linewidth}{0.1pt}\\[-2.5\baselineskip]
\begin{align}
\shortintertext{\blue{\textbf{Given}} $\gnet, \allapps, \reqs$,~\blue{\textbf{minimize:}}}
\label{eq:cost}
    \cost{\Xreq} &= \smashoperator{\sum_{\mathclap{v \in \gnet}}} \size{v} \cost{v} + 
    \smashoperator{\sum_{\mathclap{\lnet \in \gnet}}} \size{vw} \cost{vw}\phantom{ ...}\mathclap{+ \Psi}
\shortintertext{\blue{\textbf{where:}}}
\label{eq:nsize} 
\size{v} &\triangleq \sum_{\req \in \reqs} \dreq\!\!\!\! \sum_{\gapp[t] \in \treq} \sum_{i \in \mathrlap{\gapp}} \xreq<t>{i}{v} \dapp{i} \dnet{i}{v}
& \mathllap{\blue{\substack{\forall v \in \gnet}}} 
\\\label{eq:lsize} 
\size{vw} &\triangleq \sum_{\req \in \reqs} \dreq\!\!\!\! \sum_{\gapp[t] \in \treq} \sum_{\lapp \in \mathrlap{\gapp[t]}} \xreq<t>{ij}{vw} \dapp{ij} \dnet{ij}{vw}
& \mathllap{\blue{\substack{\forall \lnet \in \gnet}}} 
\\\label{eq:reject} 
\Psi &\triangleq \sum\nolimits_{\req \in \reqs} \left( 1 - \sum\nolimits_{\gapp[t] \in \treq} \xreq<t>{\user}{\oreq} \right) \dreq \psi
\shortintertext{\blue{\textbf{such that}} $\Xreq$ \blue{\textbf{satisfies} $\forall \req \in \reqs$:}}
\label{eq:onealt} 
1 &\ge \sum\nolimits_{\gapp[t] \in \treq}\xreq<t>{\user}{\oreq}
& \phantom{}\mathllap{\blue{\substack{\forall \req = ( \oreq, \areq, \dreq) \in \reqs}}} 
\\\label{eq:validroot} 
\xreq<t>{\user}{v} &= 0
&\mathllap{\blue{\substack{\forall v \neq \oreq}}}
\\\label{eq:validpathnew} 
\xreq<t>{j}{v} &= \xreq<t>{i}{v} + \sum_{\mathclap{\lnet[uv] \in \gnet}} \xreq<t>{ij}{uv} - \sum_{\mathclap{\lnet \in \gnet}} \xreq<t>{ij}{vw}
&\mathllap{\blue{\substack{\forall v \in \gnet\hfill\\\forall \lapp \in \gapp}}} 
\\\label{eq:ncapacity} 
\capacity{v} &\ge \size{v}
& \mathllap{\blue{\substack{\forall v \in \gnet}}} 
\\\label{eq:lcapacity} 
\capacity{vw} &\ge \size{vw}
& \mathllap{\blue{\substack{\forall \lnet \in \gnet}}} 
\end{align}
\rule[\baselineskip]{\linewidth}{0.4pt}\vspace{-\baselineskip}
\blue{\scriptsize
$\xreq<t>{i}{v},\xreq<t>{j}{v},\xreq<t>{ij}{uv},\xreq<t>{ij}{vw} \in \set{0,1}$}
\caption{MILP formulation for \vneap}\label{fig:vnea}
\end{figure}

\Cref{eq:onealt} reflects the fundamental difference between \vnep and \vneap. It ensures that at most one alternative is embedded for each request; it can be $0$ if the request is rejected. The request's virtual root node, $\user$, is placed at $\oreq$;
\cref{eq:validroot} requires that the root node cannot be placed anywhere else, but at $\oreq$ (\ie the \dc where this request is made). 
\ignore{\Cref{eq:validpath} ensures that $\pnet=\set{ \lnet : \xreq<t>{ij}{vw} = 1}$ is a valid (contiguous) substrate path; it requires that if there is a delta between the ingress $\lapp$ traffic into $v$ and the egress $\lapp$ traffic out of $v$, then virtual node $j$ must be on $v$.
\Cref{eq:validflow} preserves application flow; for \emph{every} virtual link $\lapp$, if virtual node $i$ is embedded on substrate $v$ then either there must be egress $\lapp$ traffic from $v$ on some substrate link $\lnet$ or $j$ must 
be located on $v$ as well.}
\Cref{eq:validpathnew} preserves application flow and ensures that $\pnet=\set{ \lnet : \xreq<t>{ij}{vw} = 1}$ is a valid (contiguous) substrate path. For \emph{every} virtual link $\lapp$, if virtual node $j$ is embedded on substrate node $v$ then either virtual node $i$ is on $v$ as well, or there is positive net flow into $v$ (ingress minus egress traffic).

The load induced by an embedding of a virtual network element of $r$ is the product of the demand $\dreq$, the element size $\dapp{}$ and the (in)efficiency $\dnet{}{}$.
\Cref{eq:nsize,eq:lsize} provide the induced load on each substrate element, by summing over all requests. The induced load must not exceed the substrate capacity limitations, as required by \cref{eq:ncapacity,eq:lcapacity}.

\Cref{eq:reject} computes the rejection cost penalty $\Psi$, which prevents optimization from minimizing costs by rejecting all requests. Rejected requests are identified using \cref{eq:onealt}. 
The coefficient $\psi$ is large enough to make rejecting a request more costly than embedding it, so requests are only rejected when capacity constraints cannot be met.
Finally, the cost optimization goal, \cref{eq:cost}, is obtained by multiplying the induced loads by their corresponding cost of resources, summing over all substrate elements, and adding the rejection penalty $\Psi$. 

\section{Solving \VNEAP}\label{sec:solution}

\ignore{We first discuss possible approaches for solving \cref{prob:vnea} and their implications. We then describe our scalable and near-optimal algorithm, \cref{tanto}, with an analysis of its properties.}

Exact solutions to either \vnep or \vneap do not scale when solving their respective MILP formulations directly. 
Hence, we propose two heuristic solutions,  \Greedy and \TANTO.

\ignore{
In this section, we discuss solutions to \cref{prob:vnea}. We start from discussing possible approaches to solving \cref{prob:vnea} in \cref{subsec:other}. It is important to remind, that the main focus of this paper is investigating whether \cref{prob:vnea} is superior to \vnep in the sense that it allows to better exploit the underlying physical substrate. As explained in \cref{subsec:other}, this question can be investigated by just comparing fractional solutions to \vnep (single alternative) and \cref{prob:vnea}.

However, this will not equip us with a practical solution to \cref{prob:vnea}. To that end, we provide a very scalable and near-optimal algorithm \cref{tanto} that provides an integral solution in \cref{sec:tanto}.
}
\subsection{GREEDY Algorithm}\label{subsec:other}

\begin{algorithm}
\caption{\textsc{Greedy}}\label{alg:greedy}\label[greedy]{greedy}
\Small
\begin{algorithmic}[1]
    \Require{$\gnet, \allapps, \reqs$}
    \Ensure{An embedding $\Xreq$ that is a feasible solution to \cref{prob:vnea}}
    \Statex \vspace{0.3\baselineskip}\hrule
    \For{$\req = (v, a, d) \in \reqs$} 
        \For {$\gapp[t] \in \tapp$}  \Comment{for each alternative virtual network topology}
            \State $\Xreq[\req]^t \gets$ greedy embedding of $\gapp[t]$ from $v$ with demand $d$ or \textit{reject}
            \Statex\Comment{Minimal cost embedding, similar to MINV}
        \EndFor
        \State $\Xreq[\req] \gets \argmin_{\Xreq[\req]^t}\set{\cost{\Xreq[\req]^t}}$ 
        \Comment{Select alternative with minimal cost}
    \EndFor
    \State \Return $\Xreq$
\end{algorithmic}
\end{algorithm}

\Greedy (\cref{alg:greedy}) uses the MINV algorithm~\cite{lin2022sft,lin2018dag} as a building block. 
MINV solves \vnep by greedily embedding each request at a minimal cost. 
For each request, \Greedy uses MINV to find an embedding for each topology alternative. It then selects the cheapest embedding.

\Greedy is a simple and easy-to-implement strategy. On the downside, it might exhaust scarce resources by preferring the lower-cost alternative as long as it can be used. This might prevent \Greedy from fully utilizing all available alternatives. 

\subsection{TANTO Algorithm}\label{sec:tanto}
\begin{table}
    \caption{Notations}
    \label{tab:notations2}
    \centering\setlength{\tabcolsep}{2pt}
    \scriptsize
    \begin{tabularx}{\columnwidth}{@{}lX@{}}
        \toprule
        \textbf{Notation} & \textbf{Description} \\ 
        \midrule
        $\Xreq$ & integral embedding \\
        $\rXreq, \rXreq[va], \rXreq[\req]$ & fractional embedding of $\reqs, \reqs[va], \req$ (relaxed \cref{prob:vnea}) \\
        $\rreqs$ & aggregated requests \\
        $\yreq<t>{i}{v}, \yreq<t>{ij}{vw} \in \Yreq$ & fractional embedding  of $\rreqs$ \\
        $\Psi(\Xreq), \Psi(\Yreq)$ & cost of rejection for $\Xreq, \Yreq$ \\
        $\WRS{\alpha(o_1), \alpha(o_2), \ldots}$ & weighted random selection from $\set{o_1, o_2, \ldots}$, using $\Vec{\alpha}$\\
        \bottomrule
    \end{tabularx}
    \vspace{-3mm}
\end{table}

In the following, we present \cref{tanto} (\cref{alg:tanto}), our scalable, near-optimal solution to \cref{prob:vnea}. \cref{tanto} is a global optimization algorithm that was adapted to \cref{prob:vnea} from the \vnep solution in~\cite{behravesh2024practical}. First (\cref{subsec:relaxation}), \cref{tanto} transforms a given instance of \cref{prob:vnea} into an LP formulation, in which embedding requests for the same application, originating at the same \dc, are aggregated together, thus, making a relaxed problem akin to a minimal cost splittable flow. This reduces the problem size and makes it independent of the number of application embedding requests. Then (\cref{subsec:rounding}), the aggregate fractional solution is randomly rounded to an integral one, allocating individual requests. 

\begin{algorithm}
\caption{\textsc{Tanto}}\label{alg:tanto}\label[tanto]{tanto}
\Small
\begin{algorithmic}[1]
    \Require{$\gnet, \allapps, \reqs$}
    \Ensure{Non-fractional embedding $\Xreq$ that is a near-optimal solution to \cref{prob:vnea}}
    \Statex \vspace{0.3\baselineskip}\hrule
    \State $\rreqs \gets \set{(v, a, \rdreq)}_{v \in \gnet, a \in \allapps}$
    \Comment{Aggregate similar requests (\cref{eq:rr})}
    \State $\Yreq \gets$ LP relaxation of \cref{prob:vnea}$(\rreqs)$ \Comment{Relaxtion (\cref{subsec:relaxation})}
    \For{$\req \in \reqs$} 
        \Comment{Rounding (\cref{subsec:rounding})}
        \State $\Xreq[\req] \gets$ \cref{embed}($\req, \Yreq)$ \Comment{$\Xreq[\req] = 0$ if $\req$ is rejected, see \cref{alg:embed}}
    \EndFor
    \State \Return $\Xreq$
\end{algorithmic}
\end{algorithm}
\vspace{-3mm}

\subsection{Aggregation and Relaxation}\label{subsec:relaxation}
 Let $\rXreq$ denote a \emph{fractional} embedding, obtained by relaxing \cref{prob:vnea} to an LP; namely, by setting $\xreq<t>{q}{s} \in [0,1]$ for every embedding of virtual network element $q$ on substrate element $s$.
 Consider the set $\reqs[va] \subseteq\reqs$ of all requests that ask for the same application at the same node, \ie $\reqs[va] = \set{\req : v=\oreq \land a=\areq}$, and let $\rXreq[va]$ denote the fractional embedding for the aggregate request $\reqs[va]$. We observe that any two embeddings $\rXreq[\req_1], \rXreq[\req_2] \in \rXreq[va]$ are \emph{swappable}, in the sense that a fraction of the traffic from $\req_1$ can be allocated using the allocation of $\req_2$, provided that the same amount of traffic from $\req_2$ is allocated using the embedding of $\req_1$. Following this observation, we define a relaxed, aggregated form of $\reqs$:
\begin{equation}\label{eq:rr}
\rreqs = \set{(v, a, \rdreq)}_{a \in \allapps, v \in \gnet} 
\text{, where } \rdreq = \smashoperator{\sum_{r \in \reqs[va]}} \dreq
\end{equation}
Let $\Yreq$ denote a relaxed solution to \vneap$(\rreqs)$, \ie when applied to $\rreqs$ instead of \reqs. \ignore{DB: was: $r$. }
Given $\Yreq$, we can readily derive a fractional solution $\rXreq$ of \vneap$(\reqs)$ and vice versa.
\begin{align}
\shortintertext{From $\Yreq$ to $\rXreq$:}
&\substack{\forall r \in \reqs,  \rreq = (\oreq, \areq, \rdreq)}:&
\begin{split}
\rxreq<t>{i}{v} &\gets \yreq<t>{i}{v} \cdot \dreq / \rdreq \\
\rxreq<t>{ij}{vw} &\gets \yreq<t>{ij}{vw} \cdot \dreq / \rdreq
\end{split}\label{eq:split}\\
\shortintertext{From $\rXreq$ to $\Yreq$}
&\substack{\forall \rreq = (v, a, \rdreq)} :&
\begin{split}
\yreq<t>{i}{v} &\gets \sum\nolimits_{r \in \reqs[va]}\rxreq<t>{i}{v} \\
\yreq<t>{ij}{vw} &\gets \sum\nolimits_{r \in \reqs[va]}\rxreq<t>{ij}{vw}
\end{split}\label{eq:merge}
\end{align}
Using \cref{eq:split}, if $\Yreq$ satisfies \crefrange{eq:onealt}{eq:lcapacity} then so does $\rXreq$, therefore if $\Yreq$ is optimal then so is $\rXreq$. Similarly, using \cref{eq:merge}, if $\rXreq$ is optimal then so is $\Yreq$. Both $\Yreq$, $\rXreq$ are solutions to an LP that can readily be obtained with a solver. Obviously, $\Yreq$ has much fewer variables than $\Xreq$, as instead of having a variable per request in $\reqs$, there is a variable per request in $\rreqs$. The size $\abs{\smash{\rreqs}}$ is bounded by all possible combinations of $v$ and $a$, but is independent of the number of application requests. 

\subsection{Rounding and Embedding of Individual Requests}\label{subsec:rounding}
We use randomized rounding to obtain an integral solution to \cref{prob:vnea}. However, we use $\Yreq$ rather than $\rXreq$. Instead of rounding each $\rxreq<t>{q}{s}$ to $\xreq<t>{q}{s}$, we round $\Yreq$ to an \emph{integral number} of requests. 

The embedding $\Yreq[va]$ can be seen as a load balancing (LB) decision on the incoming requests, defining what fraction of the overall demand of $\reqs[va]$ uses each alternative topology and what fraction of its demand is embedded on each substrate element. We can ``implement'' a fractional embedding $\Yreq[va]$ by inserting LB elements into the substrate network to split aggregate request traffic according to $\Yreq[va]$. That is, each request $\req \in \reqs[va]$ would be randomly assigned to one of its alternatives, as defined by the fractional weights $\set{\yreq<t>{\user}{\oreq}}_t$ and then randomly routed at each substrate node, according to the weights $\set{\yreq<t>{\smash{j}}{v}} \cup \set{\yreq<t>{i\smash{j}}{vw}}_{\lnet}$. 

We denote by $\WRS{\alpha(o_1), \alpha(o_2), \ldots}$ the weighted random sampling of the set $\set{o_1, o_2, \ldots}$ using the weights $\set{\alpha(o_1), \alpha(o_2), \ldots}$. 
We use $\WRS{\alpha}$ as a shorthand for $\WRS{\alpha(\textsc{true}), \alpha(\textsc{false})}$, using the weights $\alpha(\textsc{true}) = \alpha$ and $\alpha(\textsc{false}) = 1-\alpha$. 

\begin{algorithm}\caption{Embed}\label{alg:embed}\label[embed]{embed}
\Small
\begin{algorithmic}[1]
    \Require{$\req = (\oreq, \areq, \dreq), \Yreq$}
    \Ensure{Embedding $\Xreq[r]$ or $\req$ is \textbf{rejected}}
    \Statex \vspace{0.3\baselineskip}\hrule
    \State \Call{Reset}{$\Xreq[r]$}, $\rreq = (\oreq, \areq, \rdreq) \in \rreqs$ \Comment{Initialize}\label{l:init} 
    \State $d \gets \dreq / \rdreq$ \Comment{Normalized size of $\req$} \label{l:normalize}
    \State $n \gets \sum\nolimits_{\gapp[t] \in \treq} \yreq<t>{\user}{\oreq}$ \Comment{Residual capacity of $\Yreq[\rreq]$} \label{l:cap}
    \State $t \gets $\Shortstack{$\WRS{\smash{\set{\alpha(t) \text{, s.t. } \gapp[t] \in T(a(r))}}}$,  \\where $\alpha(t) = \yreq<t>{\user}{\oreq}/n $}~\phantom{MMM} \Comment{Select an alternative} \label{l:alt}
    \State \Call{Embed-or-Reject}{$\req, d, \user, \oreq$}
    \For{$\lapp \in \gapp[t]$ in topological sort order} \Comment{starting from link $\lapp[\user j]$}
        \State $v \gets w$, s.t. $\xreq<t>{i}{w} = 1$ \label{l:vi}
        \While{$\xreq<t>{j}{v} = 0 $}
            \State $n \gets \yreq<t>{j}{v} + \sum\nolimits_{\lnet \in \gnet} \yreq<t>{ij}{vw}$
            \If{$\WRS{\yreq<t>{j}{v}/n}$} \label{l:vj}
                \State \Call{Embed-or-Reject}{$\req, d, j, v$}
            \Else
                \State $n \gets n - \yreq<t>{j}{v}$
                \State $w \gets \WRS{\set{\alpha(w) \text{, s.t. } \lnet \in \gnet}}$; $\alpha(w) = \yreq<t>{ij}{vw}/n$ \label{l:ij}
                \State \Call{Embed-or-Reject}{$\req, d, \lapp, \lnet$}
                \State $v \gets w$
            \EndIf
        \EndWhile           
    \EndFor
    \Statex \vspace{0.3\baselineskip}\hrule
    \Function{Reset}{$\Xreq[r]$}   
        \ForAll{$\gapp[t]\in\treq, i, \lapp \in \gapp, v, \lnet \in \gnet$}
            \State $\xreq<t>{i}{v} \gets 0$, $\xreq<t>{ij}{vw} \gets 0$ 
        \EndFor
    \EndFunction
    \Statex \vspace{0.3\baselineskip}\hrule
    \Function{\embedreject}{$\req, d, q, s$}
        \If{$d \le \yreq<t>{q}{s}$} \label{l:fits}
            \State $\xreq<t>{q}{s} \gets 1$ \Comment{Embed} \label{l:embed}
            \State $\yreq<t>{q}{s} \gets \yreq<t>{q}{s} - d$ \Comment{Update residual capacity} \label{l:update}
        \Else  \Comment{No capacity on $\yreq<t>{q}{s}$, zero and undo embedding}
            \State $\yreq<t>{q}{s} \gets 0$ \label{l:zero}
            \For{$\xreq<t>{i}{v}$, s.t., $\xreq<t>{i}{v} > 0$} $\yreq<t>{i}{v} \gets \yreq<t>{i}{v} +d$\EndFor \label{l:restorev}
            \For{$\xreq<t>{ij}{vw}$, s.t., $ \xreq<t>{i}{v} > 0$} $\yreq<t>{ij}{vw} \gets \yreq<t>{i}{v} +d$\EndFor \label{l:restorel}
            \State \Call{Reset}{$\Xreq[r]$}
            \State $\req$ is \textbf{rejected}; abort \cref{embed}, and \Return{$\Xreq[r]$}
        \EndIf
    \EndFunction
\end{algorithmic}
\end{algorithm}
\Cref{alg:embed} finds, for each $\req \in \reqs$, an \emph{integral} embedding $\Xreq[\req]$, using the \emph{fractionally} optimal embedding $\Yreq$, obtained by the LP, as defined in \cref{subsec:relaxation}. $\Xreq[\req]$ is initialized to zero (\cref{l:init}) and remains so if $\req$ is rejected. Otherwise, $\Yreq$ is updated to reflect its new residual resource capacity after allocating $\Xreq[\req]$. One can think of embedding $\req$ using $\Yreq$ as selecting a corresponding subtree of $\Yreq$. Since $\Yreq$ is a feasible fractionally optimal solution of \cref{prob:vnea}, every allocation of $\req$ into $\Yreq$ will also be feasible as long as the residual capacity of $\Yreq$ allows embedding. 
The amount of load induced by $\rreq \in \rreqs$ on each substrate network element is specified by $\Yreq$ as fractions of $\rdreq$. We need to specify $\dreq$ in the same terms; the normalized request size $d$ is computed in \cref{l:normalize}. 

\Cref{alg:embed} first selects one of the alternative topologies $\gapp[t] \in \tapp$ (\crefrange{l:cap}{l:alt}); $t$ is chosen using a weighted random selection using the \emph{normalized} weights $\set{\yreq<t>{\user}{\oreq}}_t$. The normalization factor $n$ (\cref{l:cap}) is the available residual capacity for $\rreq$, as a fraction of its size. 
Function~\textproc{\embedreject} attempts to embed a virtual network element, $q$, onto a substrate element, $s$, inducing a normalized load $d$. If the load $d$ is less than the fractional embedding $\yreq<t>{q}{s}$ then this is a valid element embedding for $\req$; $\xreq<t>{q}{s}$ is set to $1$ (\cref{l:embed}) and $d$ is subtracted from $\yreq<t>{q}{s}$ to update the residual capacity (\cref{l:update}). However, if there is insufficient capacity, the request is rejected. This is a rounding error that means we cannot embed the full size of the request. 
The variable $\yreq<t>{q}{s}$ is zeroed (\cref{l:zero}), to prevent any more similar embedding attempts. $\Yreq$ is restored (\crefrange{l:restorev}{l:restorel}) and $\Xreq[\req]$ is reset.

\Cref{alg:embed} calls Function~\textproc{\embedreject} to embed the virtual root node of the chosen alternative, $\gapp[t]$, on the desired substrate node $\oreq$. The links and nodes of the virtual topology are then embedded in a greedy manner. The virtual topology links are examined from the root node in topological sort order (pre-order). A link $\lapp$ can only be embedded \emph{after} node $i$ is already embedded on some substrate node $v$ (\cref{l:vi}). \cref{alg:embed} first tries to embed $j$ on $v$, with the normalized probability for $\yreq<t>{j}{v}$ (\cref{l:vj}). Otherwise, it chooses one of the substrate neighbors, $w$ (again with a weighted random sampling defined by the $\Yreq[va]$, \cref{l:ij}); then tries to embed the logical link on $\lnet$; and finally updates $v$ to $w$. The process is repeated until $j$ is embedded (or $\req$ is rejected). After $j$ is embedded, \cref{alg:embed} continues with the next virtual link. 

Note that \cref{alg:embed} is amenable to parallelization, because a fractional solution for each aggregated request $\rreqs(v)$ is independent of all other aggregated requests.

\begin{figure*}
    \centering
    \begin{minipage}[b]{0.33\linewidth}%
    \centering
    \subfloat[][Main topology, {$\gapp[1]$}\label{fig:cctv_sa1}]{%
        \includegraphics[page=8]{fig/FluidModelTikz.pdf}
    }
    
    \subfloat[Alternative topology, {$\gapp[2]$}\label{fig:cctv_sa2}]{%
        \includegraphics[page=9]{fig/FluidModelTikz.pdf}
    }
    \caption{CCTV virtual network. The numbers indicate function and link sizes.}
    \label{fig:cctv}
    \end{minipage}
    \hfill
    \begin{minipage}[b]{0.6\linewidth}%
    \centering
    \setkeys{Gin}{width=0.29\linewidth}
    \subfloat[Amres-RS\\(25 nodes, 24 links)]{
        \centering
        \includegraphics[keepaspectratio=true,trim=1.5cm 1cm 1.1cm 1cm,clip]{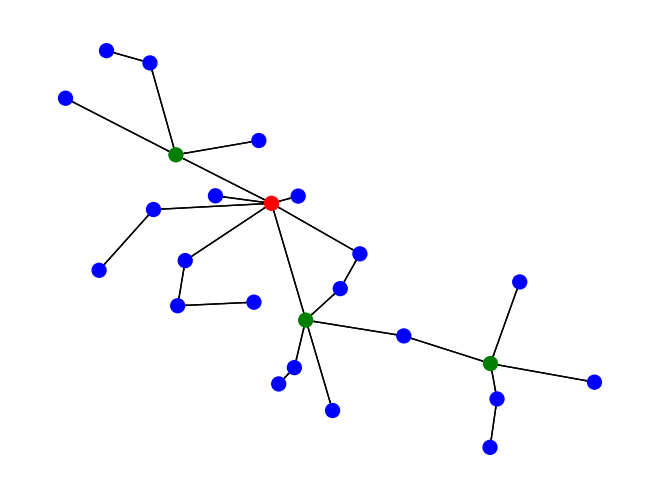}
    }\hfill
    \subfloat[Arnes-SI\\(34 nodes, 46 links)]{
        \centering
        \includegraphics[keepaspectratio=true,trim=1.5cm 1cm 0.9cm 1cm]{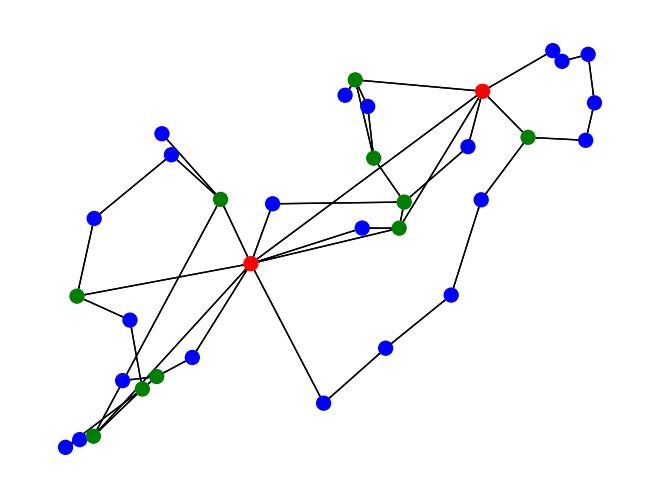}
    }\hfil
    \subfloat[Dfn-DE\\(58 nodes, 87 links)]{
        \centering
        \includegraphics[keepaspectratio=true,trim=1.5cm 1cm 1.5cm 1cm]{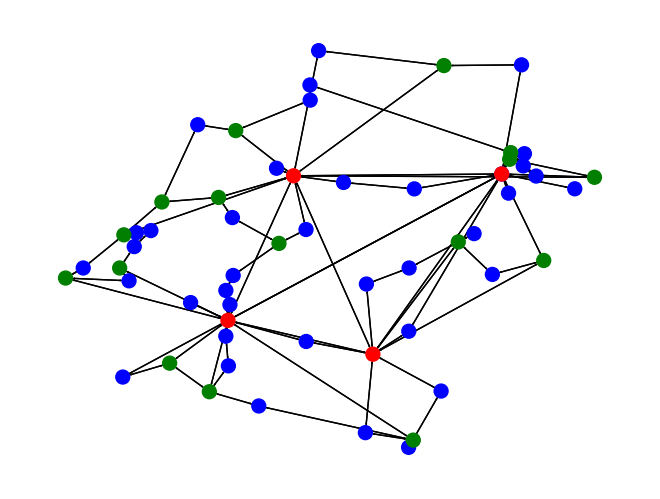}
    }
    \caption{Real network topologies. Edge, transport, and core \dcs are represented by blue, green, and red, respectively.}
    \label{fig:topologies}
    \end{minipage}%
    \vspace{-3mm}
\end{figure*}
    


\subsection{Analysis}

Solving the LP formulation of \cref{prob:vnea} yields a solution known as the ``LP fractional optimum''. 
We term this solution \OPT and note that it is a lower bound on the number of rejected requests and total embedding cost. In this subsection, we analyze the upper bound on the gap between the cost of solutions yielded by \cref{tanto} and \OPT and investigate the complexity of \cref{tanto}.


Let $\rXopt$, $\Xtant$ denote the fractionally optimal solution and the integer solution given by \cref{tanto}, respectively. Let $\Psi(\rXopt), \Psi(\Xtant)$ denote their respective rejection costs (as given by \cref{eq:reject}). 
\begin{lemma}\label{lem:reject}
    The cost of rejection added by \cref{tanto} is bounded by $\Psi(\Xtant) -\Psi(\rXopt) \le \psi\dreq^{\max}\abs{\vnet}\abs{\gnet}\abs{\Tapp}$, where $\dreq^{\max}$ denotes the maximal size of any single request and $\Tapp = \bigcup_{\app \in \allapps}\bigcup_{\app[t] \in \app}\gapp[t]$.
\end{lemma}
\begin{IEEEproof}
    Let $\Yreq$ be the aggregate flow computed in the first step of \cref{tanto} and used by the second step of \cref{tanto}. Each time a call to \cref{embed} rejects an application request, some $\yreq<t>{q}{s}$ of $\Yreq$ is zeroed (at \cref{l:zero}). Therefore, the number of rejected application requests due to rounding is bounded by the number of (non-zero) variables in $\Yreq$.
    For each $\req, t$ the number of variables $\yreq<t>{\cdot}{\cdot}$ is less than $\abs{\gnet}\abs{\gapp[t]}$. Summing over all combination of $v, \app$ and $t$, we get 
    \begin{equation*}\label{eq:bound}
    \abs{\vnet}\abs{\gnet}\sum_{\app \in \allapps}\sum_{\gapp[t] \in \tapp} \abs{\gapp[t]} =  \abs{\vnet}\abs{\gnet}\abs{\Tapp}.
    \end{equation*}
    Since $\psi\dreq^{\max}$ is the maximal rejection cost of any single request, the desired bound applies to 
    $\Psi(\Xtant) -\Psi(\Yreq)$. 
    $\rXopt$ can be aggregated using \cref{eq:merge}, and \cref{tanto} uses a fractionally optimal $\Yreq$ (\ie $\Yreq$ yields the lower bound on the number of rejected requests). Therefore, we must have $\Psi(\Yreq) = \Psi(\rXopt)$, which implies $\Psi(\Xtant) -\Psi(\rXopt) = \Psi(\Xtant) -\Psi(\Yreq)$. 
    The result follows.
\end{IEEEproof}

Let $\Xopt$ denote the optimal (integer) solution to \cref{prob:vnea}, and $\Psi(\Xopt)$ denote its corresponding rejection cost.
\begin{theorem}\cref{tanto} obtains a solution that satisfies the following deterministic bound 
 (which is independent of $\abs{\reqs}$):
    $$\Psi(\Xtant) - \Psi(\Xopt) \le \psi\dreq^{\max}\abs{\vnet}\abs{\gnet}\abs{\Tapp}$$
\end{theorem}
\begin{IEEEproof}
Since $\Psi(\Xopt) \ge \Psi(\rXopt)$, the result follows from \cref{lem:reject}.
\end{IEEEproof}
\begin{lemma}\label{lem:runtime}
    The execution time of \cref{embed} for $\req$ is of order $O(\abs{\vnet}\cdot\max_{\gapp[t] \in T(a(r))}\abs{\gapp[t]})$.
\end{lemma}
\begin{IEEEproof}
$t$ is chosen in $O(1)$. The main loop of the algorithm examines each virtual topology element in $\gapp[t]$. Each iteration examines at most $\vnet$ substrate nodes. 
\end{IEEEproof}
From \cref{lem:runtime} we get:
\begin{theorem}
The total execution time for all calls to \cref{embed} by \cref{tanto} is of order $O(\abs{\reqs}\abs{\vnet}\cdot\max_{\gapp[t] \in \Tapp}\abs{\gapp[t]})$.
\end{theorem}



\section{Evaluation}
\label{sec:evaluation}

\newcommand{\figseven}
{\begin{figure*}
\captionsetup[subfloat]{justification=centering,format=hang,captionskip=1pt}%
    \centering
    \includegraphics[scale=0.35,trim=0.3cm 0.3cm 8.5cm 0.3cm,clip]{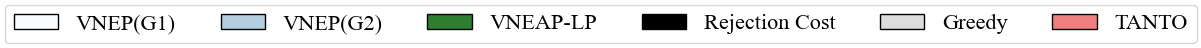}\quad\qquad\qquad
    \includegraphics[scale=0.35,trim=10cm 1.1cm 0.3cm 0.3cm,clip]{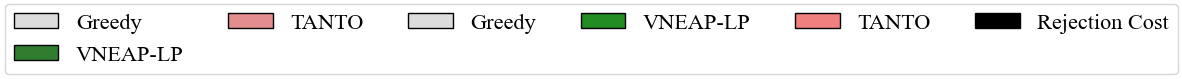}
    \\[-.7\baselineskip]\hfill
    \subfloat[Rejected demand rate]{
        \centering
        \includegraphics[width=0.22\linewidth,trim=0 0.3cm 0 0.3cm,clip]{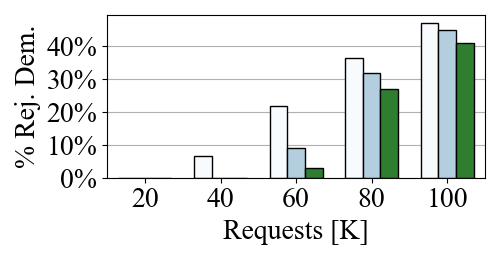}
        \label{fig:arnes_rejected_rate_opt}
    }\hfill
    \subfloat[Cost]{
        \centering
        \includegraphics[width=0.22\linewidth,trim=0 0.3cm 0 0.3cm,clip]{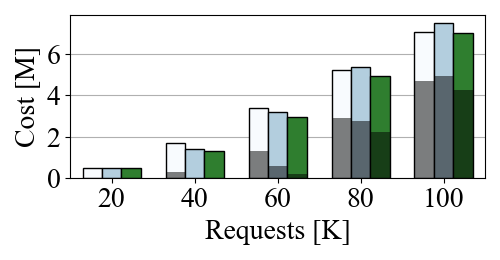}
        \label{fig:arnes_cost_opt}
    }\hfill
    \subfloat[Rejected demand rate]{
        \centering
        \includegraphics[width=0.22\linewidth,trim=0 0.3cm 0 0.3cm,clip]{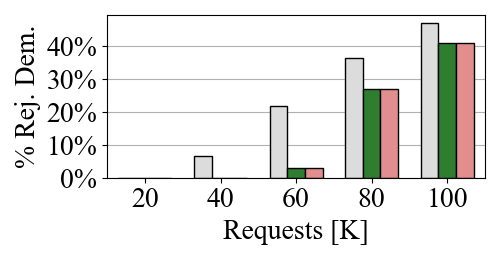}
        \label{fig:arnes_rejected_rate_tanto}
    }\hfill
    \subfloat[Cost]{
        \centering
        \includegraphics[width=0.22\linewidth,trim=0 0.3cm 0 0.3cm,clip]{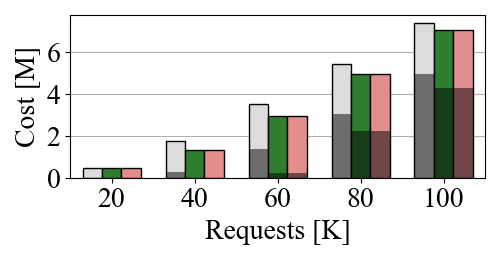}
        \label{fig:arnes_cost_tanto}
    }
    \caption{\Arnes, \NTU and \LTU equal 100\% for 60K requests.}
    \label{fig:arnes_nrf_erf1}
\vspace{-3mm}
\end{figure*}}

\newcommand{\figeightnine}
{\begin{figure*}
\captionsetup[subfloat]{justification=centering,format=hang}
\hfill%
\noindent\begin{minipage}[b]{0.47\linewidth} %
    \centering%
    \subfloat[][Two alternatives $\tapp\!=\!\set{\gapp[1],\gapp[2]}$]{
        \includegraphics[width=0.45\linewidth,trim=.4cm .4cm .3cm .3cm,clip]{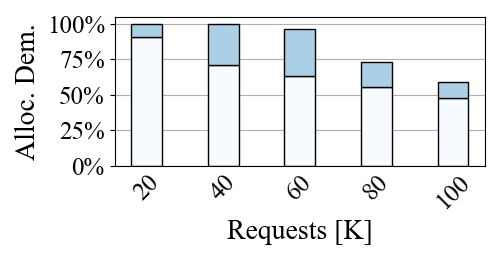}
        \label{fig:arnes_alt_share_two_opt}
    }%
    \subfloat[][Four alternatives $\tapp = \set{\gapp[1]..\gapp[4]}$]{
        \includegraphics[width=0.45\linewidth,trim=.4cm .4cm .3cm .3cm,clip]{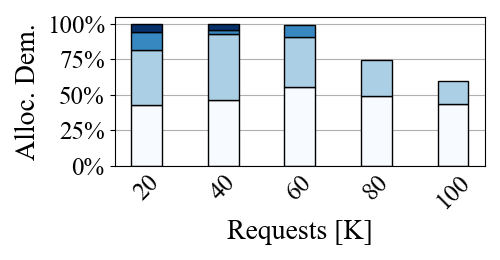}
        \label{fig:arnes_alt_share_four_opt}
    }
    \caption{\OPT: Allocated demand with each alternative in \Arnes, \NTU and \LTU equal 100\% for 60K requests.}
    \label{fig:arnes_alt_share_opt}
\end{minipage}\hfill%
\noindent\begin{minipage}[b]{0.05\linewidth}%
    \centering%
    \raisebox{\height+\baselineskip}
    {\includegraphics[width=\linewidth,trim=6cm 2cm 2.9cm 0cm,clip]{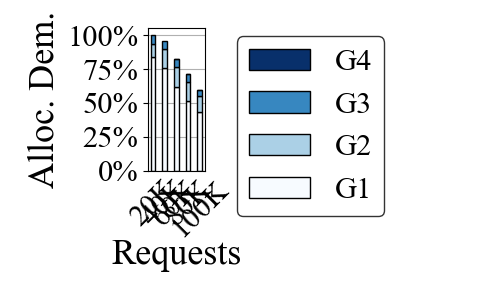}}
\end{minipage}\hfill%
\noindent\begin{minipage}[b]{0.47\linewidth}%
    \centering%
    \subfloat[Two alternatives, $\tapp = \set{\gapp[1], \gapp[2]}$]{
        \includegraphics[width=0.45\linewidth,trim=.4cm .4cm .3cm .3cm,clip]{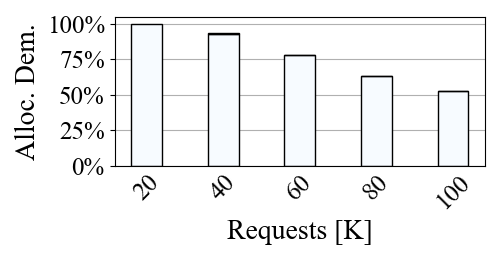}
        \label{fig:arnes_alt_share_two_greedy}
    }%
    \subfloat[Four alternatives, $\tapp = \set{\gapp[1]..\gapp[4]}$]{
        \includegraphics[width=0.45\linewidth,trim=.4cm .4cm .3cm .3cm,clip]{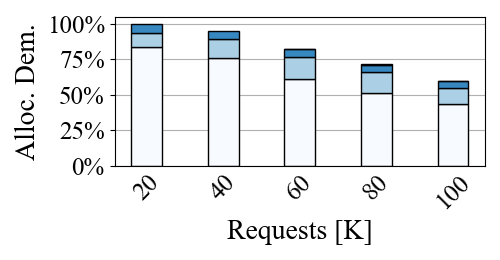}
        \label{fig:arnes_alt_share_four_greedy}
    }
    \caption{\Greedy: Allocated demand with each alternative in \Arnes, \NTU and \LTU equal 100\% for 60K requests.} 
    \label{fig:arnes_alt_share_greedy}
\end{minipage}\hfill%
\vspace{-3mm}
\end{figure*}}

We designed our experimental evaluation with three objectives in mind. Our primary objective is to understand the effect of topology alternatives on the system’s cost and utilization. 
Thus, we compare the fractional optimum \OPT, to the fractional optimum of \vnep. We obtain the latter by running the LP with a single alternative. 
Our secondary objective is to explore the integrality gap of solutions obtained by heuristic approaches. To that end, we compare \OPT to the solutions obtained by \TANTO and \Greedy. Our final objective is to understand scenarios in which global optimization is superior to the greedy approach. To that end, we compare \Greedy and \TANTO using the same alternatives.

\subsection{Experimental Setup}
\T{Virtual network} We used the CCTV application~\cite{forti2022probabilistic}, where a camera continuously captures video footage and transmits it to storage and analytics functions. This is an interesting use case, as the authors utilize an accelerator function that improves video data delivery efficiency, thus creating a tradeoff. While the function decreases the data size, it also requires additional processing resources.
We study the benefit of combining both alternatives in the context of \cref{prob:vnea}, rather than using only one of them as in \vnep.
\Cref{fig:cctv} shows two alternatives topologies $\gapp[1]$ and $\gapp[2]$. $\theta$ is a fictitious root function anchoring the request placement. 
Both topologies include four functions $f1$--$f4$, corresponding to the system driver, feature extraction, storage, and video analytics functions. The alternative topology $\gapp[2]$ adds a 
WAN accelerator function ($acc$) that reduces the traffic to $f3$.
The total function size is 105 for $\gapp[1]$ and 115 for $\gapp[2]$. Due to lack of space, we focus on this use case. However, we evaluated our solution also for other virtual topologies including trees, and obtained similar results.

\T{Substrate network} We use eight topologies from \cite{zoo}, which also include the three topologies reported in~\cite{mao2022joint} in the context of edge-to-cloud application embedding (see \cref{tab:runtime}).
To the best of our knowledge, no previous \vnep evaluation study considered larger networks. 
In alignment with the ETSI standard for multi-access edge computing (MEC)~\cite{isg2019multi,chiang2020virtual}, we classified each topology's \dcs and links into three tiers: edge, transport, and core. \Cref{fig:topologies} depicts the topologies from~\cite{mao2022joint} and \dc tier classification. Similarly to~\cite{chiang2020virtual}, we set the ratio of $3$ between the costs and capacities of \dcs and links of  successive tiers. 
We experimented with a wide range of classification and sizing options, but, due to a lack of space, we only show a subset of the results to provide sufficient insight into \OPT, \Greedy, and \TANTO. 


\T{Application requests}
\ignore{In all experiments except for the run time analysis, each algorithm processed $100K$ randomly selected application requests from a pool of $3M$ requests. In the run time analysis, each algorithm processed up to $1M$ requests from a pool of up to $30M$ requests.}
Requests originate at the edge \dcs. The request sizes follow a normal distribution, $\mathcal{N}(10,4)$. 
We derive request locations using two distributions: uniform and log-normal. The log-normal distribution creates hotspots of application requests, reflecting real-life user density in urban areas~\cite{lee2014spatial}. We utilize the uniform distribution to account for user activity in low-density regions. We ensured user density at each location is such that the overall load does not exceed the capacity of the \dc and its outgoing links.

\figseven
\figeightnine
\T{Algorithms} We used the following algorithms in our evaluation. \Greedy and \TANTO are our heuristics described in~\cref{sec:solution}. \OPT is the fractionally optimal solution of the LP described in \cref{sec:tanto}. \aOPT and \bOPT are the single-alternative fractionally optimal solutions for each alternative. 
\ignore{\Greedy is a straightforward extension of MINV~\cite{lin2022sft,lin2018dag} described in \cref{subsec:other}.} 
\ignore{It compares the embedding cost to all \dcs for each alternative and selects the most cost-effective alternative for each request, \ie opting for the lower-cost alternative when sufficient capacity is available. Our preliminary evaluation indicated that considering all possible embeddings for each alternative is infeasible. Thus, \Greedy only considers embeddings of multiple functions on the same \dc. Specifically, when embedding $\gapp[1]$, all its functions are collocated on the target \dc. When embedding $\gapp[2]$, the WAN accelerator is collocated with $f1$ and $f2$, and the remaining functions are collocated on the target \dc.} 

All algorithms were implemented in Python, utilizing CPLEX 22.1~\cite{CPLEX} to solve the LP relaxation of the MILP for \OPT. All executions were performed on Ubuntu 18.04 server equipped with an Intel Xeon Platinum 8276 CPU running at 2.1 GHz and 256GB of RAM. Each data point is obtained by averaging $30$ repetitions to obtain mean and variance with statistical fidelity. 

\T{Controlled utilization}
Following~\cite{rost2019virtual}, we define \emph{target utilization} as a ratio between the total resources required for all application embedding requests and the physical substrate resources. \ignore{Rost and Schmid~\cite{rost2019virtual} propose }
To differentiate between the node and link target utilizations in the experiments, we define \emph{node target utilization (\NTU)} and \emph{link target utilization (\LTU)} for balancing overall user demand and available capacity. 
We examined several \NTU/\LTU combinations. In all experiments except for the run-time analysis, we calibrated \NTU/\LTU using $60K$ requests. 
For example, for 100\% \NTU and \LTU, we adjusted the substrate link and \dc capacities to equal the total demand of $60K$ users requesting $\gapp[1]$, which is treated as a main topological alternative. We scaled \dc and link capacities so that the edge-transport-core capacity ratios remained consistent throughout our evaluation.

\ignore{
All algorithms were implemented in Python, utilizing CPLEX 22.1~\cite{CPLEX} to solve the LP relaxation of the MILP for \OPT. All executions were performed on Ubuntu 18.04 server equipped with an Intel Xeon Platinum 8276 CPU running at 2.1 GHz and 256GB of RAM.}

\subsection{Results}

\newcommand{\figteneleven}
{\begin{figure*}
    \captionsetup[subfloat]{justification=centering,format=hang}
    \centering
  \begin{minipage}[b]{0.59\linewidth}%
    \centering%
    \includegraphics[scale=0.32,trim=8cm 3.7cm 0.2cm 0,clip]
    {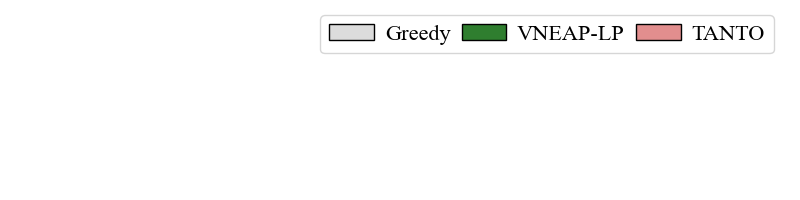}
    \mbox{}\\[-.8\baselineskip]
    \subfloat[][node-TU=100\%, link-TU=100\%\label{fig:60k_1_1}]{%
        \centering%
        \includegraphics[scale=0.25,trim=0.2cm 0 0.1cm 0cm,clip]
        {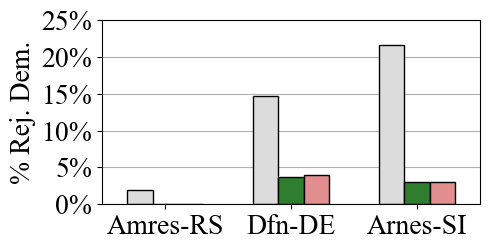}}%
    \hfill%
    \subfloat[][node-TU=50\%, link-TU=100\%\label{fig:60k_0.5_1}]{%
        \centering%
        \includegraphics[scale=0.25,trim=0.2cm 0 0.1cm 0cm,clip]
        {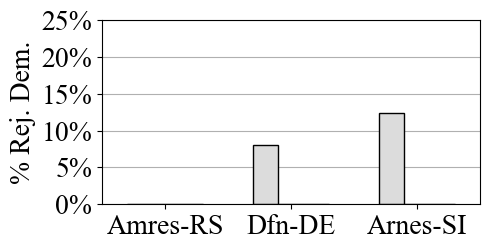}}%
    \hfill%
    \subfloat[][node-TU=100\%, link-TU=50\%\label{fig:60k_1_0.5}]{%
        \centering%
        \includegraphics[scale=0.25,trim=0.2cm 0 0.1cm 0cm,clip]
        {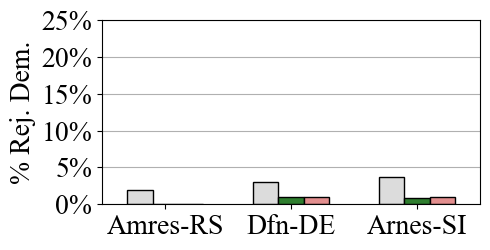}}
    \caption{Rejection rate per \NTU/\LTU combinations for three topologies (60K requests). \label{fig:60k_all}}
  \end{minipage}
  \hfill
    \begin{minipage}[b]{0.39\linewidth}%
\centering%
    \includegraphics[scale=0.25,trim=0cm 1.8cm 0.2cm 0,clip]
    {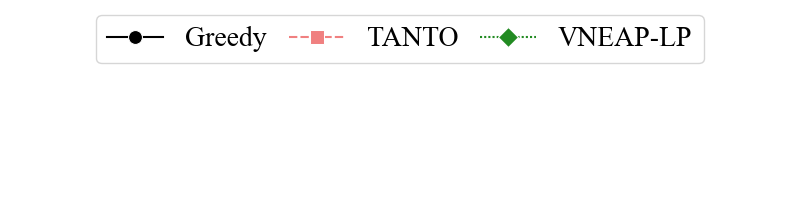}
    \mbox{}\\[-.8\baselineskip]
    \raisebox{3mm}{\includegraphics[scale=0.3,trim=0 0.2cm 0 0.25cm,clip]
        {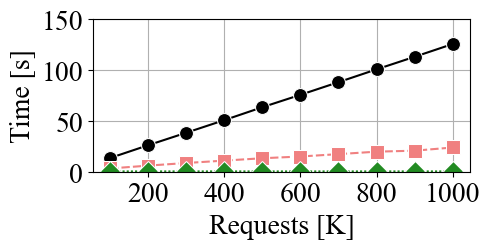}}
        \caption{Run time with CCTV virtual network on \Arnes.}
        \label{fig:arnes_runtime_cctv}
  \end{minipage}
\vspace{-5mm}
\end{figure*}}

\T{The expressive power of \vneap} 
\Cref{fig:arnes_rejected_rate_opt} presents the rejected demand rate of \OPT and the single-alternative fractionally optimal solutions \aOPT and \bOPT. 

With $20K$ requests, the rejection rate was zero, since the physical network substrate resources were sufficient to accommodate all requests. With $40K$ requests, only \aOPT rejected some requests. \OPT and \bOPT use the WAN acceleration function and had a zero rejection rate. 
At 100\% load (\ie $60K$ requests, when \NTU and \LTU equaled 100\%), all algorithms rejected some of the demand. \aOPT and \bOPT rejected 22\% and 9\% of the demand, respectively. \OPT (the fractional optimum) had a 3\% rejection rate. 
At even higher load levels, differences between each of the alternatives and \OPT were less significant because the load level actually exceeded the capacity of the network and there were fewer opportunities for any type of optimization. Nevertheless, \OPT still displayed lower rejection rates than each of the alternatives. This experiment clearly shows that the topology alternatives are useful in the global optimization, because they help to minimize the fractional optimum.  

\Cref{fig:arnes_cost_opt} shows the total embedding cost in the same experiment, with the blackened bars representing the total cost due to rejected requests (see \cref{eq:reject}). A lower rejection cost is preferable, as it indicates serving more of the demand. To account for the rejection cost consistently, $\psi$ (the lower bound on the cost of a single request rejection) was set to reflect the cost of collocating all functions of the main alternative ($\gapp[1]$) on an edge \dc, which is the most expensive embedding. Note that \OPT simultaneously served more requests and also consistently achieved the lowest total cost of embedding and the lowest rejection cost in all settings. This non-trivial result shows the power of topology alternatives when adapting to the bottlenecks in the network substrate. 


\Cref{fig:arnes_alt_share_two_opt} shows the percentage of allocated demand achieved by \OPT, broken down by alternatives ($\gapp[1]$ and $\gapp[2]$). With $20K$ requests, the abundant capacity led \OPT to choose the more cost-effective alternative, $\gapp[1]$. With $40K$ and $60K$ requests, some links became fully saturated, leading to increased use of $\gapp[2]$. 
\OPT traded the cost of capacity incurred by  $\gapp[2]$ (additional capacity for the $acc$ function) for rejection rate minimization. At even higher loads, there were fewer opportunities to apply acceleration, leading to a decreased usage of $\gapp[2]$ by \OPT. 

\begin{figure}
    \centering
\includegraphics[width=0.9\linewidth,trim=.2cm 0.3cm 0.3cm 0.2cm,clip]{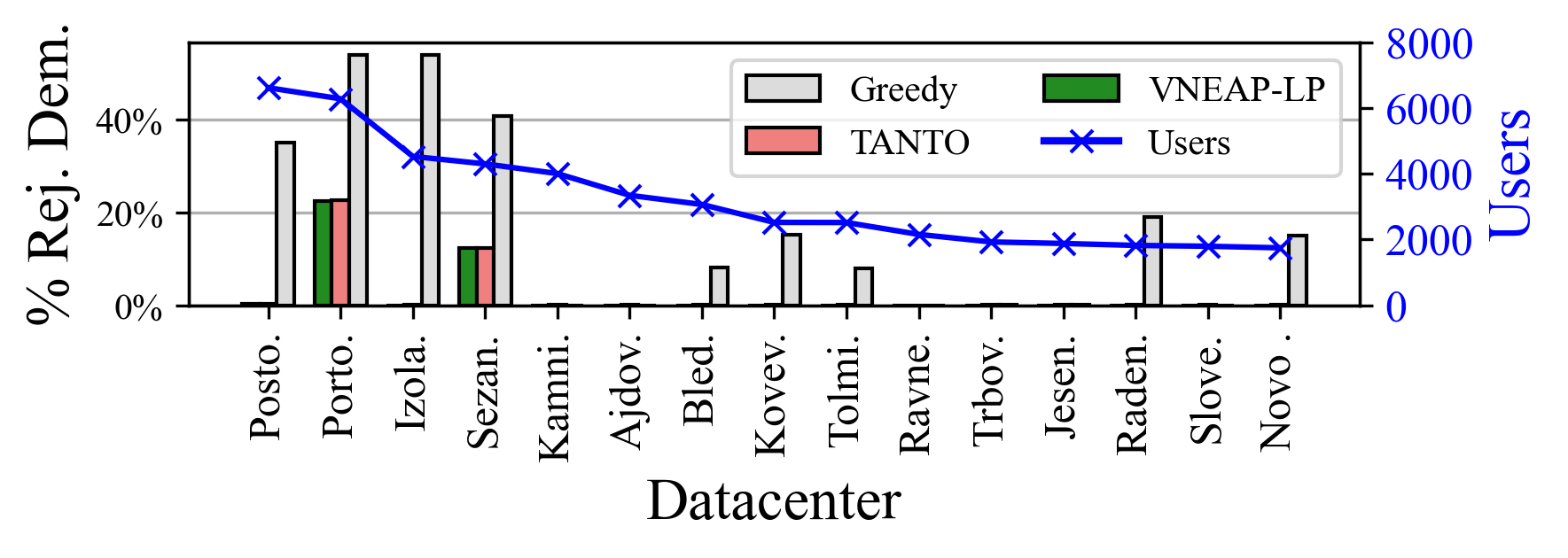}
    \caption{User count and rejection rate in first 15 \dcs in \Arnes for \NTU and \LTU of 100\% with 60K users.}\label{fig:arnes_60k}
    \vspace{-3mm}
\end{figure}

\figteneleven

\Cref{fig:arnes_alt_share_four_opt} depicts the results of the same experiment with four alternatives, $\gapp[1]$--$\gapp[4]$. Similar to the CCTV virtual network, $\gapp[1]$ is the main alternative, while $\gapp[2]$--$\gapp[4]$ are accelerated alternatives. These alternatives have accelerator ($acc$) sizes of 1, 5, and 10, and traffic reduction levels of 50\%, 75\%, and 95\%, respectively. This setup highlights the tradeoff between accelerator sizes and traffic reduction levels.

\Cref{fig:arnes_alt_share_four_opt} shows the percentage of allocated demand. With $20K$ requests, \OPT served 82\% of the demand with $\gapp[1]$ and $\gapp[2]$, while $\gapp[3]$ and $\gapp[4]$ were used for 12\% and 6\%, respectively. $\gapp[3]$ and $\gapp[4]$ featured a significant traffic reduction which led to cost savings on expensive links, justifying the additional cost incurred by the WAN accelerator. As the load (number of requests) increased, the benefit of these alternatives diminished due to the reduction in available capacity. For each load level, \OPT selected the optimal combination of alternatives to maximize the number of embedded requests at the lowest possible cost. Together, the two experiments demonstrate the complex relationship between the operating conditions and the optimal embedding. At the same time, these results clearly show that \OPT outperforms the fractional solution of \vnep. This means that considering topology alternatives for virtual network embedding is beneficial.

\T{The integrality gap of heuristic solutions}
We repeat the experiment from~\cref{fig:arnes_alt_share_opt} with \Greedy. \Cref{fig:arnes_alt_share_greedy} shows the percentage of allocated demand it achieved with two and four alternatives, respectively. Unlike \OPT, \Greedy embedded a similar combination of alternative topologies in all load levels. With two alternatives (\cref{fig:arnes_alt_share_two_greedy}), it preferred the more cost-effective $\gapp[1]$ as long as there was available capacity, and used $\gapp[2]$ to satisfy only a few requests before the links became fully congested. With four alternatives (\cref{fig:arnes_alt_share_four_greedy}), \Greedy typically favored the less resource-demanding $\gapp[1]$, and used $\gapp[2]$ and $\gapp[3]$ for the paths in which they were more cost-effective. Nonetheless, \Greedy was oblivious to the total load level and used a similar ratio of $\gapp[2]$ and $\gapp[3]$ in all settings.
As a result, \Greedy allocated less demand than \OPT in most settings. In the settings where they allocated similar rates of the demand, \OPT cost was lower than that of \Greedy. For example, \OPT cost was $5\%$ lower with $20K$ requests.

\Cref{fig:arnes_rejected_rate_tanto}
shows the rejected demand rate of \TANTO, \Greedy, and \OPT and \cref{fig:arnes_cost_tanto} shows the corresponding total cost of embedding. 
As the results show, \TANTO consistently outperformed \Greedy, and the integrality gap between \TANTO and \OPT was negligible, which confirms that \TANTO is near-optimal.
As one can expect, \TANTO consistently outperformed \Greedy even at extremely high load levels (higher than 100\% utilization of the system at 80K and 100K requests). However, as one would also expect, the differences became less pronounced as the system became overloaded.

\T{Detailed comparison under non-uniform spatial demand distribution} \Cref{fig:arnes_60k} shows the rejected demand rate for \NTU and \LTU of 100\% with $60K$ requests in \Arnes, broken down by user locations. 
We ordered the \dcs in descending order of the number of users that generate requests. The figure displays the first $15$ \dcs that together serve 80\% of the entire user population.
In the first four \dcs, the rejection rate of \Greedy was at least 28\% higher than that of \OPT and \TANTO. Without global optimization, \Greedy struggled to effectively embed requests originating from the congested \dcs because of the limited link resources. In the remaining \dcs, most requests were successfully embedded even though \Greedy still exhibited a non-zero rejection rate in some cases. \OPT and \TANTO exhibited similar behavior in all \dcs.

\Cref{fig:60k_all} displays the rejected demand rate for the three network topologies from~\cite{mao2022joint}  and $60K$ requests. To understand the relative effect of link and node utilization, we study three different scenarios in which utilization levels alternate between nodes and links. 
In all experiments, \OPT and \TANTO consistently achieved similar rejection rates. In most cases \Greedy performed worse than \TANTO. One exception to occurs in \Amres: this is a a relatively small and simple topology, which limits optimization decisions of any strategy. Thus, on this topology, \Greedy is on par with \OPT and \TANTO.

The differences between \cref{fig:60k_0.5_1} and
\cref{fig:60k_1_0.5} show how the behavior of the algorithms changes with the nature of the bottlenecks in the physical substrate. When the physical links are primary bottlenecks (\cref{fig:60k_0.5_1}), we see a clear advantage of global optimization compared to \Greedy. When we switch the situation by making physical links the primary bottleneck (\cref{fig:60k_1_0.5}), \Greedy rejects less demand because it has much more spare link capacity available before it becomes saturated. Nevertheless, even though the rejection rate in \TANTO and \OPT slightly grows, global optimization is still advantageous to \Greedy.

\T{Run-time analysis} \Cref{fig:arnes_runtime_cctv} shows the mean run time of \TANTO and \Greedy, for the CCTV virtual network in \Arnes and request ranges from $100K$ to $1M$. This is two orders of magnitude larger than any previous study except~\cite{behravesh2024practical}. To eliminate the impact of load level observed in previous experiments, \NTU and \LTU were set to 100\% for each data point, leading to consistent rejection rates of $3\%$ and $21\%$ for \TANTO and \Greedy, respectively. \OPT represents the runtime of the LP phase of \TANTO. It was circa 1 second across all data points, thanks to aggregating requests originating at the same \dc. The runtime of \TANTO increased linearly from 3 seconds for $100K$ requests to $24$ seconds for $1M$ requests. In contrast, \Greedy's runtime increased up to $128$ seconds. Although the runtime of both algorithms scales linearly with the number of requests, the second phase of \TANTO can be executed in parallel (see \cref{sec:solution}), whereas \Greedy must process each request sequentially.

\ignore{We repeated this evaluation with the tree virtual network in \cref{fig:tree}. It has the same total function size as the CCTV application and features three leaves that can be accelerated. The runtime of \TANTO was similar to the CCTV experiment, and that of \Greedy increased by $13$ seconds for $1M$ requests. \TANTO was barely impacted by the complexity of the virtual network thanks to the parallel execution of its second phase.  } 


\Cref{tab:runtime} summarizes the run times and rejection rates for \TANTO and \Greedy with $1M$ requests and \NTU = \LTU = 100\%, across all our evaluated topologies. The run time of \TANTO varied from 18 to 52 seconds and was mainly influenced by topology size. \Greedy is more sensitive to the topology size: its run time was $5\times$ to $10\times$ higher than that of \TANTO.

\begin{table}
\caption{Run time and percentage of rejected demand for 1M requests with CCTV virtual network, \NTU and \LTU equal 100\%.}
\label{tab:runtime}
\centering\scriptsize\setlength{\tabcolsep}{2pt}
\begin{tabularx}{\linewidth}%
{@{}lccXcccXccc@{}}
\toprule
\multicolumn{3}{c}{\textbf{Topology}} 
&& \multicolumn{3}{c}{\textbf{Run time} [sec]} 
&& \multicolumn{2}{c}{\textbf{Rejected ratio}} 
\\\cmidrule(lr){1-3} \cmidrule(lr){5-7} \cmidrule(lr){9-10}
\textbf{Name} & \textbf{Nodes} & \textbf{Links} && \OPT &  \TANTO & \Greedy && \TANTO & \Greedy 
\\\midrule
Xspedius & 34 & 49  && 0.2 & 17.9 & 177     && 5\% & 20\% \\
\Amres & 25 & 24    && 0.1  & 18.3 & 154    && 0\% & 2\% \\
SwitchL3 & 30 & 51  && 0.3 & 20.4 & 158     && 3\% & 13\% \\
\Arnes & 34 & 46    && 0.3  & 21.2 & 127    && 3\% & 21\% \\
\Dfn & 58 & 87      && 1.3 & 25 & 170       && 3\% & 14\% \\
NetworkUSA & 35 & 39 && 0.4 & 25.9 & 172    && 1\% & 3\% \\
Telcove & 71 & 70   && 1.4 & 32.7 & 290     && 2\% & 6\% \\
TataNld & 143 & 181 && 10.5 & 51.8 & 261    && 6\% & 11\% \\
\bottomrule
\end{tabularx}
\vspace{-3mm}
\end{table}
\section{Conclusions}\label{sec:conclusions}

We presented a novel problem, \cref{prob:vnea}, which generalizes the classical \vnep---a central problem in network virtualization. \cref{prob:vnea} captures the ability to select from multiple admissible application configurations, a domain property that was not modeled before. We demonstrated the higher expressive power of this new problem through theoretical and experimental analysis. 
We proposed two highly scalable algorithms, \Greedy and \TANTO, to solve \vneap. Our analysis and evaluation on various realistic topologies demonstrated that \TANTO consistently outperforms the greedy strategy and obtains near-optimal solutions (\ie with negligible integrality gaps). Furthermore, we showed that parallelizing \TANTO significantly reduced runtime compared to the greedy approach. 
In future work, we plan to extend the model to include energy minimization and latency constraints and explore the online version of \VNEAP. While generating the set of alternative topologies is outside this work's scope, it represents an intriguing problem we intend to investigate further. Also, we plan to extend our results to general graphs by
considering graph tree decomposition as proposed in~\cite{RostDohneSchmid-parameterized-2019}.

\clearpage


{
    \balance
    \bibliographystyle{IEEEtran}
    \bibliography{IEEEabrv,0main}
}


\end{document}